\documentclass[12pt]{article}
\usepackage{amsmath}
\usepackage{amsfonts}
\usepackage{amssymb}
\usepackage{gensymb}
\usepackage{times}
\usepackage{graphicx}
\usepackage{color}
\usepackage{multirow}
\usepackage{bm}
\usepackage{array}
\newcolumntype{M}[1]{>{\centering\arraybackslash}m{#1}}
\usepackage{subcaption}
\usepackage{tikz}
\usetikzlibrary{bayesnet}
\usepackage[authoryear]{natbib}
\usepackage{rotating}
\usepackage{bm}
\usepackage{latexsym}
\renewcommand{\eqref}[1]{Eq.(\ref{#1})}
\newcommand{\EV}[1]{\mathbb{E}\left[ #1 \right]}
\newcommand\independent{\protect\mathpalette{\protect\independenT}{\perp}}
\def\independenT#1#2{\mathrel{\rlap{$#1#2$}\mkern2mu{#1#2}}}

\title{Integrating Flexible Normalization into Mid-Level Representations of Deep Convolutional Neural Networks}
\author{Luis Gonzalo Sanchez Giraldo, Odelia Schwartz\\
  Computer Science Department\\
  University of Miami\\
  Coral Gables, FL 33146 \\
    \texttt{\{lgsanchez, odelia\}@cs.miami.edu}\\
  }
  \date{}
\begin{document}
\maketitle
\begin{abstract}
  Deep convolutional neural networks (CNNs) are becoming increasingly popular models to predict neural responses in visual cortex. However, contextual effects, which are prevalent in neural processing and in perception, are not explicitly handled by current CNNs, including those used for neural prediction. In primary visual cortex, neural responses are modulated by stimuli spatially surrounding the classical receptive field in rich ways. These effects have been modeled with divisive normalization approaches, including flexible models, where spatial normalization is recruited only to the degree responses from center and surround locations are deemed statistically dependent. We propose a flexible normalization model applied to mid-level representations of deep CNNs as a tractable way to study contextual normalization mechanisms in mid-level cortical areas. This approach captures non-trivial spatial dependencies among mid-level features in CNNs, such as those present in textures and other visual stimuli, that arise from tiling high order features, geometrically. We expect that the proposed approach can make predictions about when spatial normalization might be recruited in mid-level cortical areas. We also expect this approach to be useful as part of the CNN toolkit, therefore going beyond more restrictive fixed forms of normalization.
  \end{abstract}

\section{Introduction}
It has long been argued that an important step in understanding the information processing mechanisms in the brain is to understand the nature of the input stimuli \citep{FAttneave1954, HBarlow1961}. Visual processing of natural images is a paradigmatic example that has been studied extensively \citep{ESimoncelli2001, LZhaoping2006, LZhaoping2014, BOlshausen2014, WGeisler2008, AHyvarinen2009}. Structure in images can be captured in the form of statistical dependencies among the responses of filters acting on the image at different scales, orientations, and spatial locations \citep[e.g.][]{ABell1997, BOlshausen1997, AHyvarinen2009}. These regularities often manifest in a nonlinear fashion \citep{ESimoncelli1997, BWegmann1990,CZetzsche2005,JGolden2016}. Therefore, it is natural to think that neural processing systems employ nonlinear operations to exploit dependencies, appropriately.  
\par
Both perception and neural responses are influenced by the spatial context, i.e., by stimuli that spatially surround a given point in space. Spatial contextual influences beyond the classical receptive field have been extensively documented for neurons in primary visual cortex \citep[e.g.][]{JLevitt1997,MSceniak1999,JCavanaugh2002a,JCavanaugh2002b}. Models that are based on nonlinear statistical regularities across space in images have been able to capture some of these effects \citep{RRao1999,OSchwartz2001,MSpratling2010,YKarklin2009,MZhu2013,RCoenCagli2012,TLochmann2012}.
\par
Here, we focus on divisive normalization, a nonlinear computation that has been regarded as a canonical computation in the brain \citep{DHeeger1992, MCarandini1997, MCarandini2012}. From a coding perspective, divisive normalization acts as a transformation that reduces nonlinear dependencies among filter activation patterns in natural stimuli \citep[e.g.][]{OSchwartz2001}. Different forms of divisive normalization have been considered in modeling spatial contextual interactions among cortical neurons. In its basic form, the divisive normalization operation is applied uniformly across the entire visual field. However, spatial context effects in primary visual cortex are better explained by a weighted normalization signal \citep[e.g.][]{JCavanaugh2002a, JCavanaugh2002b, OSchwartz2001}. Recently, more sophisticated models that recruit normalization in a nonuniform fashion \citep{RCoenCagli2012} have shown better generalization at predicting responses of V1 neurons to natural images \citep{RCoenCagli2015}. The rationale behind this form of \emph{flexible normalization} (and related predictive coding models of \citep{MSpratling2010,TLochmann2012}) is that contextual redundancies vary with stimulus. In the flexible normalization model, divisive normalization is therefore only recruited at points where the pool of spatial context filter responses to an image are deemed statistically dependent with the filter responses in a center location. This relates to highlighting salient information by segmentation in regions of the image in which the spatial homogeneity breaks down \citep{ZLi1999}.
\par
As basic computational modules, it would be expected that non-linearities take place at different stages of the cortical processing hierarchy. However, studying these operations beyond the primary visual cortex level, for instance understanding when normalization is recruited for natural images, has been rather difficult.
Divisive normalization models rely on access to individual neural unit responses, which are then combined to produce the modulation effect from the pool of units. In comparison to primary visual cortex, where different features such as orientation, spatial frequency and scale have a fairly well understood role in characterizing visual stimuli, the optimal stimulus space for mid-cortical levels is less well understood \citep{TPoggio2016}. 
\par
In this work, we propose the use of deep CNNs to study how flexible normalization might work at intermediate level representations. CNNs have shown intriguing ability to predict neural responses beyond primary visual cortex \citep{NKriegeskorte2015,DYamins2016,DPospisil2016,RCichy2016,NLaskar2018}. In addition, CNNs have interestingly incorporated simplified forms of normalization \citep{KJarrett2009, AKrizhevsky2012, MRen2017}.  While we don't claim this to be a direct model of V2 responses, second layer neural units do combine the responses from units with V1-like features lending to a larger repertoire of possible stimuli acting at a higher level. CNNs can therefore provide a tractable way to model representations that might be employed by mid-levels of the visual processing hierarchy. Here, we integrate flexible normalization into the AlexNet CNN architecture \citep{AKrizhevsky2012}, although our approach can be more broadly applied to other CNN and hierarchical architectures.
\par
For mid-level representations, we show that incorporating flexible normalization can capture non-trivial spatial dependencies among features such as those present in textures, and more generally, geometric arrangements of features tiling the space. Our focus here is on developing the framework for the CNN and demonstrating the learned statistics and spatial arrangements that result for middle layers of the CNN. We expect the proposed approach can make predictions about when spatial normalization might be recruited in mid-level areas, and therefore will be useful for interplay with future neuroscience experiments as well as become a standard component in the CNN toolkit.

\subsection{Contributions of this work}
Divisive normalization is ubiquitous in the brain, but contextual surround influences in vision have mostly been studied in area V1. Numerous reasons have made difficult furthering our understanding beyond V1. Divisive normalization models rely on access to individual neural unit responses, which are then combined to produce the modulation effect from the pool of units. In V1, models such as steerable pyramids or Gabor filters provide a good account for the neural receptive field, and are often a front end to more sophisticated normalization models. However, good models of V2 responses have been more elusive, and consequently, it is unclear when normalization is recruited in area V2. Surround influences with naturalistic stimuli have only recently been addressed in neurophysiology experiments with natural texture images \citep{CZiemba2018}.
\par
CNNs have shown compatibility to neural data inside the classical receptive field, making it a good starting point. Here, we focus on the second convolutional layer of AlexNet. Second layer neural units combine V1-like features captured by the first layer units into bigger receptive fields. Since the network is constrained to capture information relevant to natural images, we expect the second layer will only learn such structures that are meaningful and not all possible combinations of simple features.
\par
The model of contextual interaction we propose to use is a normative model based on the premise that one of the purposes of normalization is to remove high order dependencies that cannot be removed by linear or point-wise non-linearities and that extend beyond the classical receptive field (This also means they extend beyond the reach of the max pooling layers). This class of model has been used to explain V1 contextual influences, but it has not been applied to higher order units \citep{RCoenCagli2009,RCoenCagli2012}. Our results for the second layer make predictions about when normalization might be recruited in area V2. We also applied our approach to the first layer units as a control.
This approach could be adopted to other hierarchical architectures and higher layers, and thus has more general applicability.
\par
From a technical standpoint, models such as the mixture of GSMs and flexible normalization have been studied extensively for V1. Our main technical contribution is making these models applicable to higher order units that do not have clear orientation structure as in V1, and demonstrating it on CNN units. This is non-trivial. For V1 filters, symmetry constraints on the covariance matrix can be assumed, given the orientation structure of the receptive fields. For higher order units, the symmetry is not obvious and cannot be assumed. We found that learning proceeds well without symmetry constraints by modifying the model of \citep{RCoenCagli2009} as described in Section \ref{sec:normalization_in_deep_CNNs}.

\section{Normalization in Deep Neural Nets}\label{sec:normalization_deep_neural_nets}
Recently, new forms of normalization have been introduced to the deep neural networks tool set \citep{SIoffe2015, LBa2016}. The motivation for these computations is different from the divisive normalization models in neuroscience, which are based on observations of neural responses. Batch normalization \citep{SIoffe2015} is a popular technique aimed at removing the covariate shift over time (i.e., in batches) in each hidden layer unit, with the goal of accelerating training by maintaining global statistics of the layer activations. Layer normalization \citep{LBa2016} on the other hand, employs averages across units in a given layer (and space in the case of convolutional networks) at every time step, introducing invariances in the network that benefit the speed of learning. Batch and layer normalization provide better conditioning of the signals and gradients that flow through the network, forward and backwards, and have been studied from this perspective. 
\par
Simple forms of divisive normalization that draw inspiration from neuroscience, such as those described in \citep{KJarrett2009,AKrizhevsky2012}, have been used to improve the accuracy of deep neural network architectures for object recognition. However, the empirical evaluation of deeper architectures in \citep{KSimonyan2015} reached a different conclusion showing that the inclusion of local response normalization (LRN) did not offer any significant gains in accuracy. One possible yet untested hypothesis for this case is that the increased depth may be able to account for some of the nonlinear behavior associated with LRN. Nevertheless, it is important to note that these empirical conclusions have only considered simple and fairly restrictive forms of normalization and measured their relevance solely in terms of classification accuracy\footnote{\normalsize{Robustness to adversarial examples could be used to evaluate the role of normalization.}}. 
\par
Recent work that attempts at unifying the different forms of normalization discussed above has started to reconsider the importance of normalization for object recognition, in the context of supervised deep networks \citep{MRen2017}. In their work, divisive normalization is defined as a localized operation in space and in features where normalization statistics are collected independently for each sample. Divisive normalization approaches arising from a generative model perspective have also been recently introduced \citep[e.g.][]{JBalle2016}. Other work on supervised networks inspired by primary visual cortex circuitry has proposed normalization as a way to learn discriminant saliency map between a target and its null class \citep{SHan2010,SHan2014}. Although these works extend beyond the simple normalization forms discussed in previous paragraphs, they are still limited to fixed normalization pools and to early stages of processing. None of these approaches have thus far considered the role of spatial dependencies and normalization in middle layers of CNNs to address questions in neuroscience. 
\par
Our work extends the class of flexible normalization models considered in \citep{RCoenCagli2012,RCoenCagli2015} which stem from a normative perspective where the division operation relates to (the inverse of) a generative model of natural stimuli. In previous work, flexible normalization models were learned for an oriented filter bank akin to primary visual cortical filters. We develop a flexible normalization model that can be applied to convolution filters in deep CNNs. Our objective in this paper is to develop the methodology and to study the statistical properties and the structure of the dependencies that emerge in middle layers (specifically, we focus on the second convolutional layer of a AlexNet). We expect our model to be useful in providing insights and plausible hypotheses about when normalization is recruited in visual cortical areas beyond Primary Visual Cortex.
The flexible model can also be used in the future for computer vision applications such as object recognition, but is beyond the scope of this paper.
          
\section{Background}
We describe the Gaussian Scale Mixture and flexible normalization model, which serves as a background to our modeling.
\subsection{Statistical model for divisive normalization}
A characteristic of natural stimuli is that the coefficients obtained by localized linear decompositions such as wavelets or independent component analysis are highly non-Gaussian, generally depicting the presence of heavy tailed distributions \citep{DField1987}. In addition, these coefficients, even if linearly uncorrelated, still expose a form of dependency where the standard deviation of one coefficient can be predicted by the magnitudes of spatially related coefficients \citep{ESimoncelli1997}. In this sense, models that extend beyond linearity are needed to deal with nonlinear dependencies that arise in natural stimuli.
\par
A conceptually simple yet powerful generative model that can capture this form of coupling is known as the Gaussian Scale Mixture \citep{DAndrews1974, MWainwright1999, MWainwright2000}. In this class of models, the multiplicative coordination between filter activations is captured by incorporating a common mixer variable where local Gaussian variables are multiplied by a common mixer. After forming the generative model, one can reduce the dependencies and estimate the local Gaussian variables, via inversion (divisive normalization). The Gaussian variables may themselves be linearly correlated which amounts to a weighted normalization.
\par
Formally, a vector $X$ containing a set of $m$ coupled activations is obtained by multiplying an independent positive scalar random variable $V$ (which we denote the mixer variable) with an $m$-dimensional Gaussian random vector $G$ with zero mean and covariance $\bm{\Lambda}$, that is, $X  = VG$. The random variable $X \vert V = v$ is a zero mean Gaussian random variable with covariance $\bm{\Lambda}v^2$, and $X$ is distributed with pdf:
\begin{equation}
p_X(\mathbf{x}) = \int\limits_{0}^{\infty} \frac{v^{-m}}{\left(2 \pi \right)^{m/2} \vert\bm{\Lambda} \vert^{1/2}}\exp{\left(-\frac{\mathbf{x}^{\mathrm{T}}\bm{\Lambda}^{-1}\mathbf{x}}{2 v^2}\right)} p_V(v)\mathrm{d}v.
\end{equation}
For analytical tractability, we consider the case where the mixer $V$ is a Rayleigh distributed random variable with pdf, $ p_V(v) = \frac{v}{h^2}\exp{\left(-\frac{v^2}{2h^2}\right)},\; \textrm{for}\; v \in \left[0, \infty\right),$ and scale parameter $h$. Integrating over $v$ yields the following pdf:
\begin{equation}\label{eq:pdf_of_gsm}
p_X(\mathbf{x}) = \frac{1}{\left(2 \pi \right)^{m/2} \vert\bm{\Lambda} \vert^{1/2}h^m} a^{1-m/2}K_{m/2-1}(a),
\end{equation}
where $K_{\lambda}(\cdot)$ is the modified Bessel function of the second kind, and
\begin{equation}\label{eq:GSM_normalization_gain}
a^2 = \frac{\mathbf{x}^{\mathrm{T}}\bm{\Lambda}^{-1}\mathbf{x}}{h^2}.
\end{equation}
To ease notation, we can let $\mathbf{\Lambda}$ absorb the scale parameter $h$. 
Reversing the above model to make inferences about $G$ given $X$ results in an operation similar to divisive normalization. Given an instance $\mathbf{x}$ of $X$, we can compute the conditional expectation of the $i$th element of $G$ as follows:
\begin{equation}\label{eq:conditional_x_given_c}
\EV{g_i|\mathbf{x}} =  \frac{x_i}{\sqrt{a}}\frac{K_{\frac{m-1}{2}}(a)}{K_{\frac{m}{2}-1}(a)}.
\end{equation}
The divisive normalization is weighted, due to the term $a$, which incorporates the inverse of the covariance matrix in the computations of the normalization factor.

\subsection{Flexible contextual normalization as a mixture of Gaussian scale mixtures}
The Gaussian Scale Mixture model described above captures the coordination between filter activations (e.g., for receptive fields that lie in nearby spatial locations) through a single mixer variable. The normalization operation produced by performing inference on the GSM model is replicated across the entire image intrinsically assumes the statistics to be homogeneous across space. However, the statistical dependency between filter activations may vary depending on the particular visual image and set of filters, such as if the filters cover a single visual object or feature, or are spaced across the border of objects in a given image \citep{OSchwartz2006,OSchwartz2009}. 
\par
A more sophisticated model \citep{RCoenCagli2009}  \citep[see also][]{RCoenCagli2012}, uses a two-component mixture of GSMs,   
\begin{equation}\label{eq:binary_gsm}
p_X(\mathbf{x}) = \Pi_{\textrm{cs}}p_X(\mathbf{x}\vert \bm{\Lambda}_{\textrm{cs}}) + (1 - \Pi_{\textrm{cs}})p_{X_{\textrm{c}}}(\mathbf{x}_{\textrm{c}}\vert \bm{\Lambda}_{\textrm{c}})p_{X_{\textrm{s}}}(\mathbf{x}_{\textrm{s}}\vert \bm{\Lambda}_{\textrm{s}}),
\end{equation}  
where $\mathbf{x}_{\textrm{c}}$ and $\mathbf{x}_{\textrm{s}}$ denote the set of responses from units with receptive fields in center and surround locations, and $\bm{\Lambda}_{\textrm{cs}}, \bm{\Lambda}_{\textrm{c}}$, and $\bm{\Lambda}_{\textrm{s}}$ are the parameters that capture covariant structure of the neural responses. In this model, normalization is only recruited to the degree center and surround responses are deemed as statistically dependent. The first term of \eqref{eq:binary_gsm}, $p_X(\mathbf{x}\vert \bm{\Lambda}_{\textrm{cs}})$, corresponds to center-surround dependent units. In the center-surround dependent component, the responses are coupled linearly by the covariance $\bm{\Lambda}_{\textrm{cs}}$ and nonlinearly by the multiplicative mixer. The product $p_{X_{\textrm{c}}}(\mathbf{x}_{\textrm{c}}\vert \bm{\Lambda}_{\textrm{c}})p_{X_{\textrm{s}}}(\mathbf{x}_{\textrm{s}}\vert \bm{\Lambda}_{\textrm{s}})$ in the second term represents the statistical independence between the center group and the surround group.

\section{Normalization in Deep Convolutional Networks}\label{sec:normalization_in_deep_CNNs}
We next describe our approach for incorporating flexible normalization into
convolutional layers of deep CNNs. We also explain how we modified the mixture of GSM
model of \citep{RCoenCagli2009} to accomodate this.

\begin{figure}[t]
\centering
\includegraphics[width=0.6\linewidth]{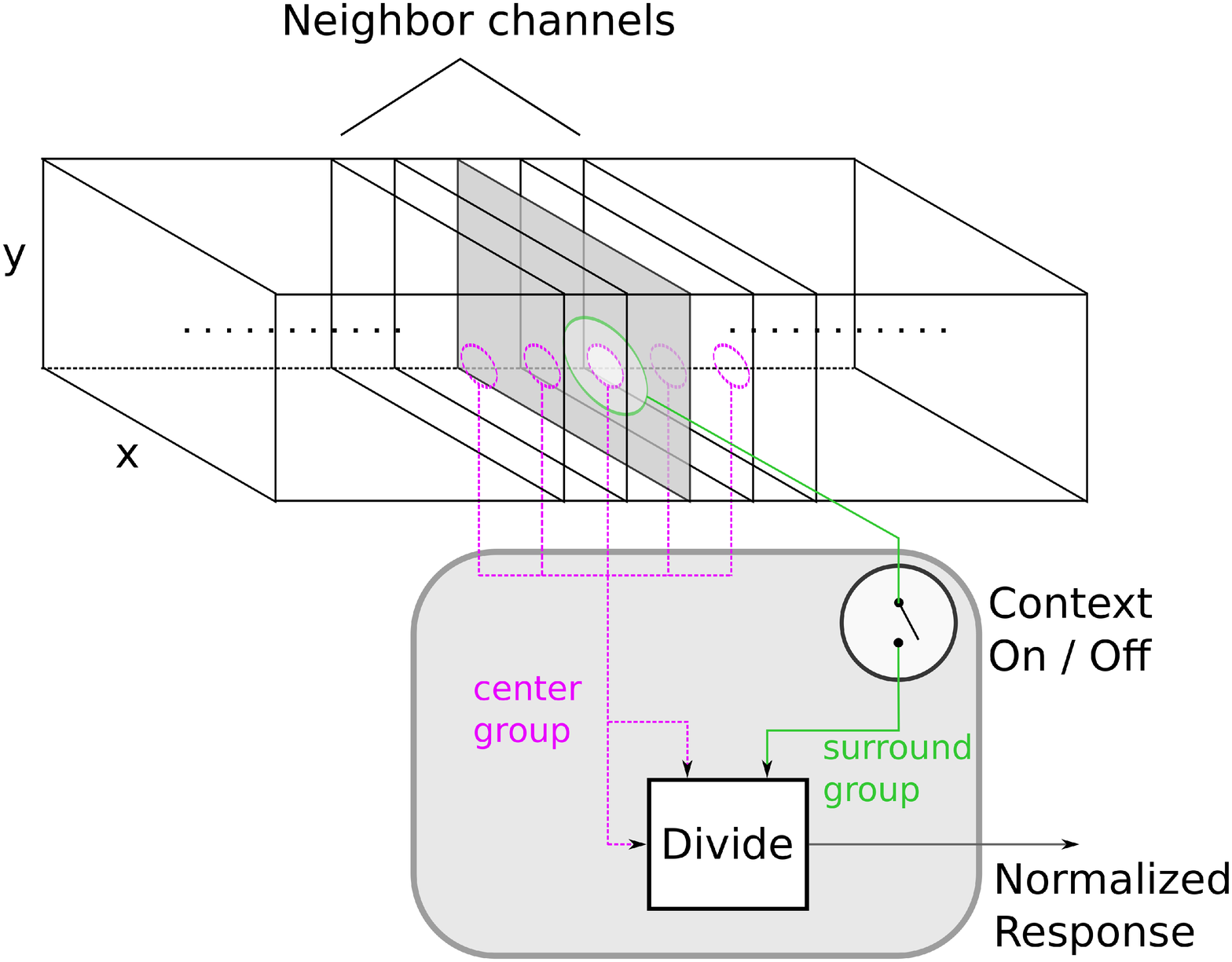}
\caption{Schematic of flexible normalization on map computed by a convolutional layer of a deep CNN. As with flexible normalization, the surround normalization is gated and determined based on
inference about statistical dependencies across space. To compute the normalized response of a filter $k$ at location $(x,y)$, the model uses responses from adjacent filters (channels) in the arrangement (akin to cross-orientation suppression in primary visual cortex) as the center group, and a set of responses from the same filter $k$ at relative displacements from the $(x,y)$ position to form the surround group (spatial context).}\label{fig:CNN_flexible_normalization}
\end{figure}

\subsection{Convolutional layers and Flexible Normalization}
In their most basic form, convolutional neural networks are a particular instance of feed forward networks where the affine component of the transformation is restricted by local connectivity and weight sharing constraints. Convolutional layers of deep CNNs are arrangements of filters that uniformly process the input over the spatial dimensions. On $2$-dimensional images, each of the CNN filters linearly transforms a collection of $2$-dimensional arrays called input channels. For instance, RGB images are $2$-dimensional images with $3$ channels. The output produced by each filter is a $2$-dimensional array of responses called a map. Therefore, each convolutional layer produces as many output maps as filters. Let $I_{\textrm{in}}(x,y,\ell)$ be the collection of $2$-d input arrays, where $x$ and $y$ denote the spatial indexes and $\ell \in \mathcal{X}_{\textrm{in}}$ the input channel index \footnote{\normalsize{It should be clear from the context when $x$ refers to spatial location and not to a realization of $X$}}. A convolutional layer is a collection of $3$-dimensional arrays $\{W_{k}(x,y,\ell)\}_{k \in \mathcal{C}_{\textrm{out}}}$. The operation of convolution, which yields a map, is defined as:
\begin{equation}\label{eq:NN_convolution}
I_{\textrm{out}}(x,y,k) = \sum\limits_{\ell}\sum\limits_{x', y'} I_{\textrm{in}}(x + x',y + y',\ell)W_{k}(x',y',\ell)
\end{equation}
\par
In addition to convolutions and point-wise nonlinearities, CNNs can include other nonlinear operations such as pooling and normalization whose outputs depend on the activities of groups of neural units. Here, we cascade the flexible normalization model with the output map of a convolution layer. Flexible normalization of the outputs of a convolutional layer is carried out at each spatial position and output channel of $I_{\textrm{out}}$. For channel $k$ and spatial position $(x, y)$, the normalization pool consists of two groups: (i) a group of activations at the same $(x,y)$ position from spatially overlapping filters from neighboring channels to $k$, called the center group. We use the center group that is already part of the Alexnet CNN local normalization layer (akin to cross-orientation suppression in V1 \citep{DHeeger1992}); (ii) a set of responses from the same filter $k$ at spatially shifted positions, called the surround group. According to the flexible normalization, the surround normalization is gated and determined based on inference about the statistical dependencies between center and surround activations.
\par
Figure \ref{fig:CNN_flexible_normalization} depicts this arrangement of maps produced by the filters in a convolutional layer as a $3$-dimensional array. For each map $k$, we compute the normalized response at each $(x, y)$ location using the flexible normalization model introduced above. 

\subsection{Flexible normalization for convolutional layers}
\begin{figure}[t]
\centering
\includegraphics[]{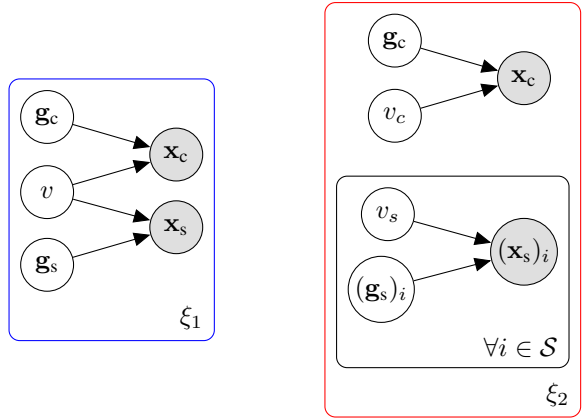}
\caption{Flexible normalization model, based on a mixture of Gaussian Scale Mixtures. (left) Center-Surround dependent; (right) Center-Surround Independent. The model is similar to \citep{RCoenCagli2009,RCoenCagli2015}, except that when center units are independent from surround units, we further impose independence of the surround unit activations. This removes the need impose any symmetry constraints in learning the model parameters, which is critical for this approach to work in higher CNN layers beyond V1-like oriented filters.}\label{fig:binary_gsm_factor_graph}
\end{figure}

One of the main differences between our model and \citep{RCoenCagli2009} is that our model imposes statistical independence among surround responses in the center-surround independent component of the mixture. This is achieved by making: 
\begin{equation}\label{eq:surround_factorization}
p_{X_{\textrm{s}}}(\mathbf{x}_{\textrm{s}}\vert \bm{\Lambda}_{\textrm{s}}) = \prod_{\ell \in \mathcal{S}}p_{\left(X_s\right)_\ell}\left(\left(\mathbf{x}_{\textrm{s}}\right)_{\ell}\right \vert \left(\bm{\Lambda}_{\textrm{s}}\right)_{\ell}).
\end{equation} 
In other words, when the center units are independent from the surround units, the group of surround units do not share the same mixer. By having independent mixers in our model, we avoid making any assumptions about symmetries in the responses of the surround units. Symmetry constraints based on the orientation of the V1 model units were originally used in \citep{RCoenCagli2009} for learning the parameters of the model. It is important to bear in mind that for mid-level representations there is no clear intuition or explicit knowledge about the nature of the symmetries that may arise across space. A graphical model of the flexible normalization model proposed here is depicted in Figure \ref{fig:binary_gsm_factor_graph}.

\subsection{Inference}
Another key difference between our model and \citep{OSchwartz2009} is the inference. In our model,  we assume there exists a common underlying Gaussian variable $\hat{G}$ that generates both types of responses (center-surround dependent and center-surround independent). The coupling is therefore, a two-stage process. First, a latent response vector $\hat{G}$ is sampled from a Gaussian distribution with zero mean, and identity covariance. This response is then linearly mapped by one of two possible transformations depending on whether the response is center-surround dependent or independent. Subsequently, the multiplicative coupling is applied to the linearly transformed vector according to the type of response (dependent or independent). The main reason for the above choice is that, if we were only resolving the multiplicative couplings, the distribution of the inferred response would  still be a mixture of Gaussians which cannot be decoupled by linear means.  
\par
Reversing the coupling by computing $\EV{\hat{G}_i\vert \mathbf{x}}$ is also a two-stage process. First, posterior probabilities of $\mathbf{x}$ being center-surround dependent are obtained using Bayes rule, $ p(\xi_{1}\vert \mathbf{x}) = \frac{p(\mathbf{x} \vert \xi_{\textrm{1}})\Pi_{\textrm{cs}}}{p(\mathbf{x})}.$
Then, conditional expectations $\EV{G_i\vert \mathbf{x}, \bm{\Lambda}_{\textrm{cs}}}$ and $\EV{G_i\vert \mathbf{x}_{\textrm{c}},\bm{\Lambda}_{\textrm{c}}}$ are linearly mapped to a common space. Namely, we apply a linear transformation $\mathbf{Q}^{\mathrm{T}}$ to the center-surround independent component $\textrm{cs}\independent$, such that:
\begin{equation}\label{eq:CS_ind_toCS_dep}
\mathbf{Q}^{\mathrm{T}}\bm{\Lambda}_{\textrm{cs}\independent}\mathbf{Q} = 
\mathbf{Q}^{\mathrm{T}}\left[\begin{array}{c|c}
\bm{\Lambda}_{\textrm{c}} & \mathbf{0} \\ \hline
\mathbf{0} & \bm{\Lambda}_{\textrm{s}}
\end{array}\right] \mathbf{Q} = \bm{\Lambda}_{\textrm{cs}},
\end{equation}
Inference in our flexible normalization model is given by:
\begin{equation}\label{eq:inference_without_whitening}
  \EV{\hat{G}_i\vert \mathbf{c}} = p(\xi_{1}\vert \mathbf{x})\EV{G \vert \mathbf{x}, \bm{\Lambda}_{\textrm{cs}}} + (1 - p(\xi_{1}\vert \mathbf{x}))\left(\mathbf{Q}^{\mathrm{T}}\right)_{i, :}\EV{G \vert \mathbf{x},\bm{\Lambda}_{\textrm{c}}, \bm{\Lambda}_{\textrm{s}}},
\end{equation}
where $(\mathbf{Q}^{\mathrm{T}})_{i,:}$ denotes the $i$th row of $\mathbf{Q}^{\mathrm{T}}$. This inference can be followed by whitening of the components of $\hat{G}$ yielding the desired identity covariance matrix, $\mathbf{I}$. However, here, the relevant operation is the transformation that takes one covariance and makes it equal to the other covariance matching the distributions of the center-surround dependent and center-surround independent component after removing the multiplicative couplings (\eqref{eq:CS_ind_toCS_dep}).

\subsubsection*{Learning parameters of the model}
In this work, our main purpose is to observe the effects of normalization in the responses obtained at the outputs of a convolutional layer in a deep CNN. For this reason, we apply the flexible normalization model to  the responses of filters from a pre-trained network that does not include flexible normalization \footnote{\normalsize{In our work, as with the original AlexNet, filters were trained for object recognition}}. The responses of a layer from these pre-trained network are used to construct the set of center and surround units to be normalized. The parameters of the flexible normalization model, the prior $\Pi_{\textrm{cs}}$ and covariances $\bm{\Lambda}_{\textrm{cs}}$, $\bm{\Lambda}_{\textrm{c}}$, and $\bm{\Lambda}_{\textrm{s}}$, are then learned by Expectation Maximization (EM) fitting to the pre-trained CNN responses \citep{RCoenCagli2009}(see appendix for details). 
\section{Simulations}
We integrate flexible normalization into the AlexNet architecture \citep{AKrizhevsky2012} pre-trained on the ImageNet ILSVRC2012 object recognition challenge. Since our main goal is to investigate what the effects of normalization are at the layer level rather than at the network level, we only learn the parameters of the divisive normalization model on top of the pre-trained filters. The divisive normalization is applied to the outputs of the convolutional layer.
In particular, we integrate flexible normalization into the outputs of the second convolutional layer of AlexNet. We also consider the first layer of AlexNet as a control.
\par
We focus on the second layer for two reasons. First, it is comprised of combinations of V1-like units in the first layer, and so is likely to be more tractable in future studies comparing to neurophysiology studies in V2. Second, we found empirically that on average, the responses of layer 3 units in AlexNet are much less statistically dependent across space, suggesting that from an efficient coding perspective divisive normalization across space would have less influence as we move up the hierarchy.
\par
In our model, the center neighborhoods are the same built-in neighborhoods that were induced by the local response normalization operation carried out in the original AlexNet architecture. The surround groups are obtained by taking activations from an approximately circular neighborhood with a radius of 4 strides apart, at every $45$ degrees, which yields a total of 8 surround units. Figure \ref{fig:normalization_neighborhood} depicts the spatial arrangement of a center response and the positions of its surround responses.

\subsection{Redundancy in activations of mid-layers}
As argued above, multiplicative couplings between linear decomposition coefficients are common in natural images. As we show below, activations at intermediate layers such as Conv2 display a significant amount of coupling.
\par
Focusing on the Conv2 layer from AlexNet, we examine the structure of spatial dependencies within a unit. 
We show that even at spatial locations for which the filters have less than $20\%$ overlap, the values of the activations of spatially shifted units expose high order correlations\footnote{\normalsize{Correlations beyond first order include correlation of squares as a special case}}.
%
\begin{figure}[t]
\centering
\begin{subfigure}[b]{0.45\linewidth}
\centering
\includegraphics[trim=0 -20 0 -20, clip, height=0.4722\linewidth]{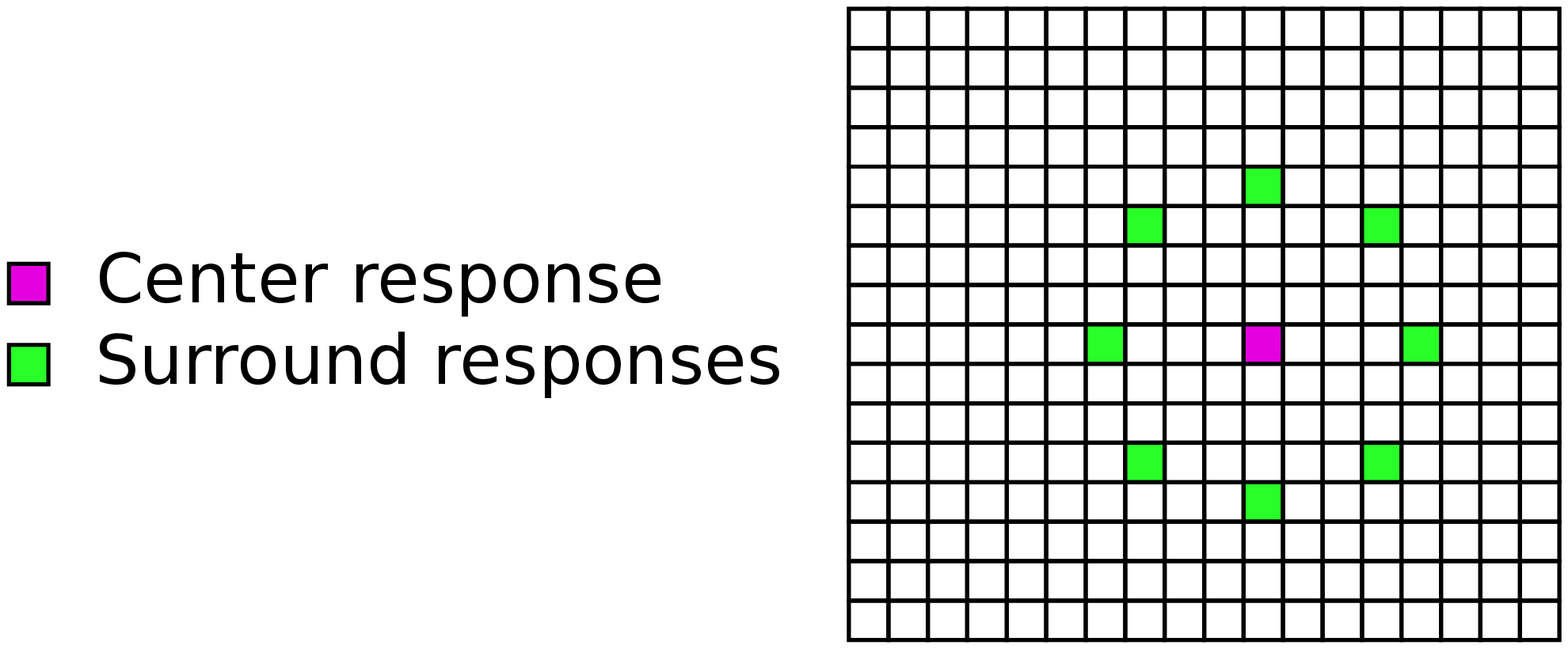}
\caption{}\label{fig:normalization_neighborhood}
\end{subfigure}
\hspace{0.025\linewidth}
\begin{subfigure}[b]{0.5\linewidth}
\centering
\includegraphics[trim=40 25 110 25, clip,height=0.425\linewidth]{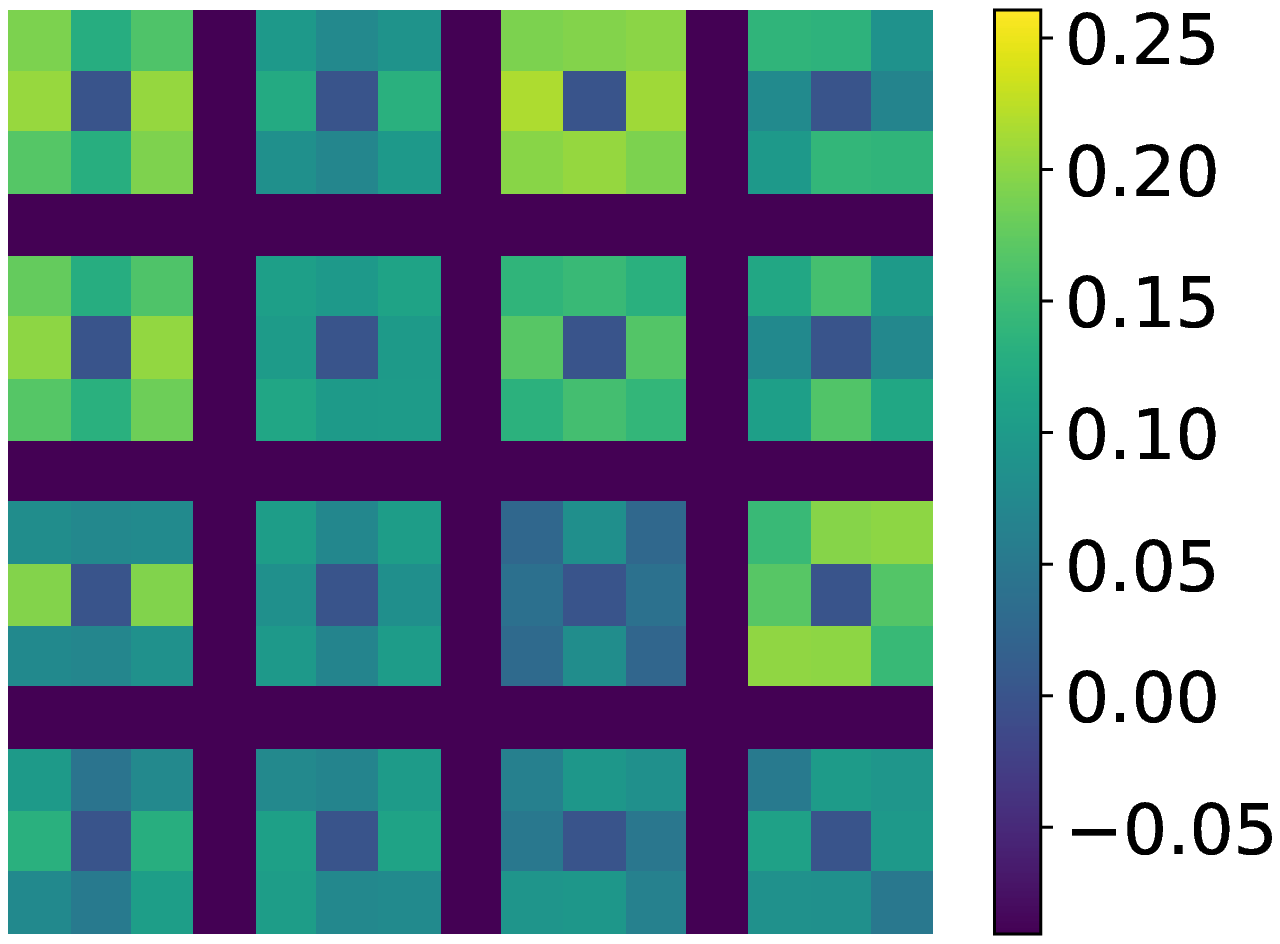}
\includegraphics[trim=70 25 20 25, clip,height=0.425\linewidth]{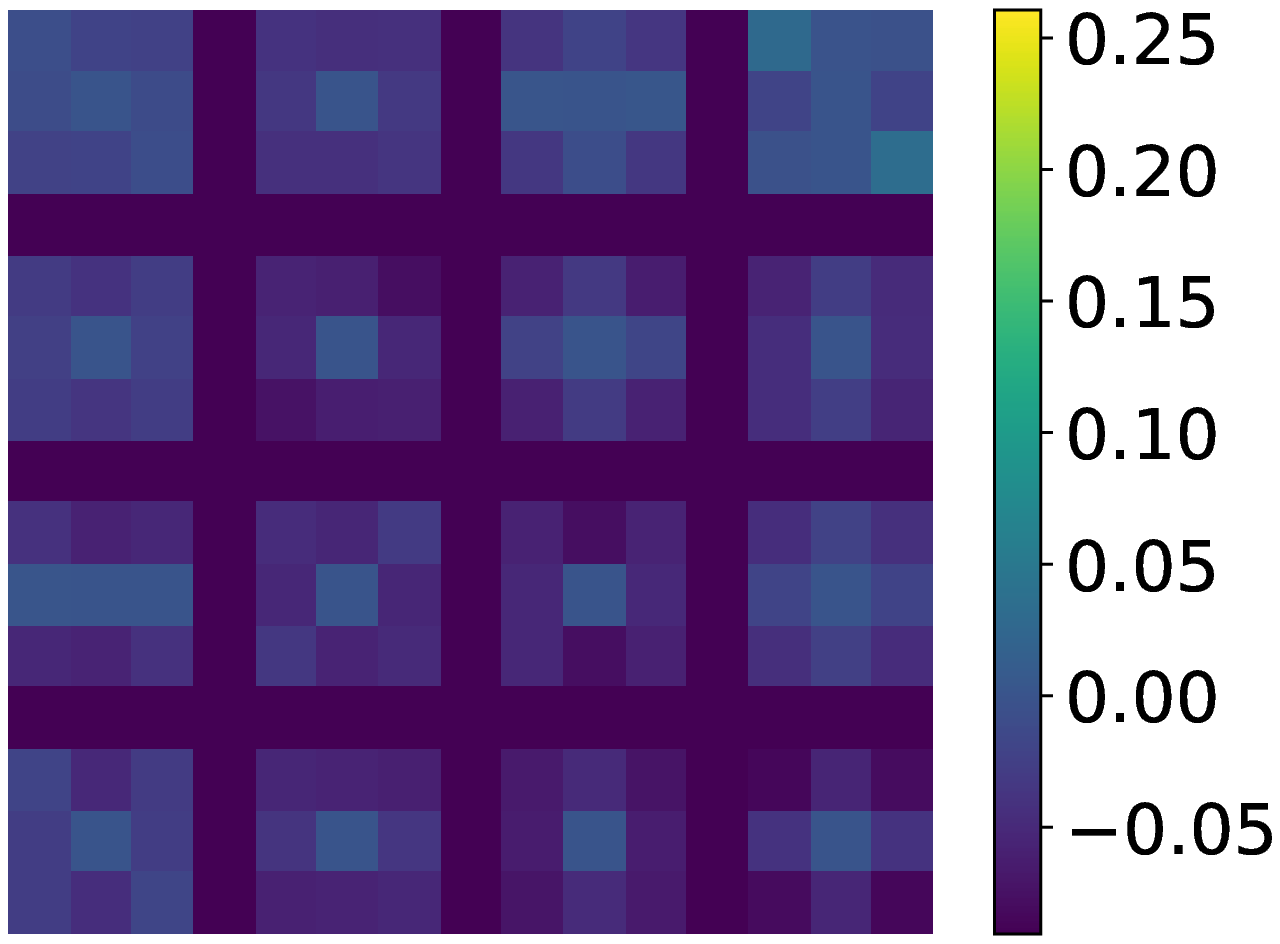}
\caption{}\label{fig:Energy_correlation_conv2_units}
\end{subfigure}
\caption{Energy correlations in the second convolutional layer of AlexNet before and after flexible normalization. (a) Spatial distribution of center and surround activations in the normalization pool. (b) Correlation of energies between center and surround responses from a subset of 16 Conv2 units from AlexNet before (left) and after (right) flexible normalization. Each of the sixteen $3 \times 3$ tiles depicts the correlation between the center activation and each of the 8 surround units shown in (a). It is clear that normalization reduces the energy correlations.}
\end{figure}
In Figure \ref{fig:Energy_correlation_conv2_units}, we display the energy correlations for the activations of a subset of units in the second convolutional layer (Conv2) of AlexNet. For each unit, we display the correlation of energies between the given unit and its spatial neighbors 4 strides apart in either the vertical or horizontal direction. Each one of the $3 \times 3$ tiles is the corresponding squared correlation for a particular unit. We see that not only do these high order couplings remain for the outputs of the Conv2 layer, but also the regularities of how their values are distributed across space. For various units, it is clear that spatial shifts in particular directions have stronger couplings.

\subsection{Dependency reduction by flexible normalization}
To visually assess the effect of normalization on the correlation structure among units, we depict the joint conditional histograms of the unit activations, after normalization and whitening. Previous studies with V1-like filters have shown that filter activations follow a bowtie-like structure that can be understood as a high order coupling \citep[e.g.][]{OSchwartz2001}. In particular, the amplitude of one variable gives information about the variance (standard deviation) of the other variable. This dependency can be reduced via divisive normalization from neighboring filter activations.
\begin{figure}[!t]
  \centering
  \includegraphics[]{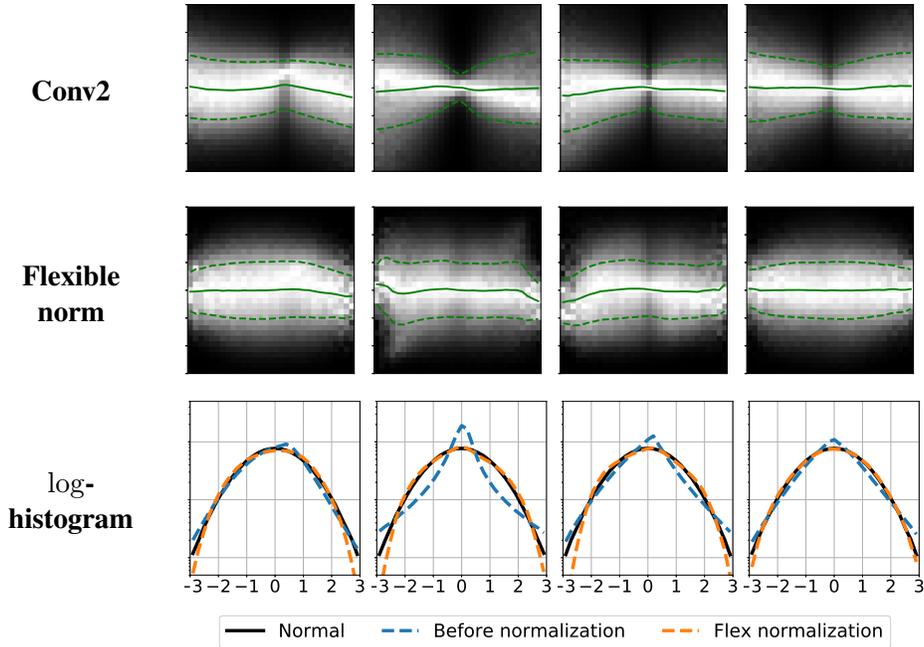}
  \caption{Marginal and joint conditional distributions of activations from exemplar Conv2 units in AlexNet, before an after flexible normalization. The joint conditional distributions are a simple way to visually inspect dependencies. As we can see, Conv2 units at different spatial locations are nonlinearly coupled. Flexible normalization reduces these dependencies, making the conditional distributions look closer to constant. In addition, the marginal $log$-histograms show that the normalized responses become closer to Gaussian, in agreement with the model assumptions.}
\label{fig:bowties_log_histograms_conv2}
\end{figure}
Figure \ref{fig:bowties_log_histograms_conv2} shows the conditional histograms ($p(x_{\textrm{s}}\vert x_{\textrm{c}})$) for the same pair of center-surround unit activations before and after applying flexible divisive normalization. Along with the normalized conditional histograms, we show marginal log-histograms which give an idea of how normalization changes the marginal distributions from highly kurtotic to more Gaussian-like.
\par
\begin{figure}[!t]
  \centering
  \includegraphics[width=0.6\linewidth]{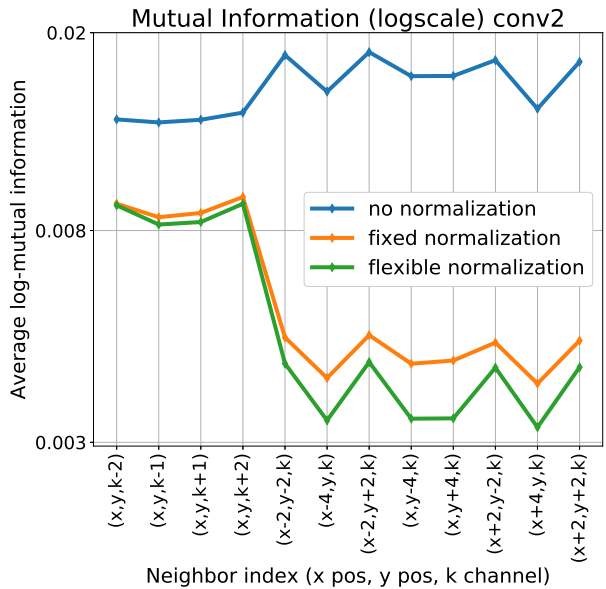}
  \caption{Population mutual information summary statistics for flexible normalization versus the control surround normalization model. Average over channels of the mutual information (x axis) in $\log$ scale between unit $(x, y, k)$ and its relative neighbors.}\label{fig:mutual_information_conv2}
  \end{figure}

We further quantify the results for the flexible normalization model, and compare to a simpler baseline surround normalization model using a single GSM. 
At a population level, both normalization models, flexible and single GSM, consistently reduce mutual information between the center unit and spatial surround activations ($2032$ out of all $8 \times 256$ center-surround pairs). But the flexible normalization reduces mutual information beyond the level achieved by the control model (Figure \ref{fig:mutual_information_conv2}).
The distribution of the difference of the estimates of mutual information between single GSM minus flexible normalization is skewed to the right ($1.8$ skewness).
\par
\begin{figure}[!t]
  \centering
  \includegraphics[width=0.7\linewidth]{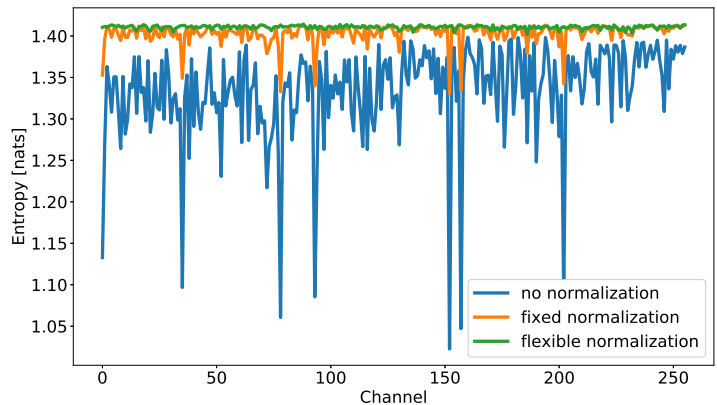}
  \caption{Population entropy summary statistics for flexible normalization versus the control surround normalization model. Marginal entropies from standardized responses (zero mean and unit variance) before and after normalization for each unit in the second convolutional layer of AlexNet.}\label{fig:entropy_conv2}
  \end{figure}

Also, computing the entropy of the activations before and after normalization shows consistent increase which is more pronounced in the flexible normalization model (Figure \ref{fig:entropy_conv2}). Since the activations before and after normalization have been scaled to have unit variance, larger entropies correspond to random variables whose distributions are more similar to the Gaussian. We have also examined the expected likelihood of the test data. For more than 98\% of the units, flexible normalization has higher likelihood compared to the single GSM model. Overall, the population quantities confirm that flexible normalization is better than the single GSM at capturing the Gaussian statistics and reducing the statistical dependencies.    

\subsection{Predicting homogeneity of stimuli based on mid-level features}
The main idea of flexible normalization is that contextual modulation of neural responses should be present only when responses are deemed as statistically dependent. In the case of V1, co-linearity of stimuli in the preferred direction of the center unit would cause the flexible normalization model to invoke suppression (see also appendix for Conv1 units). In other words, the model would infer high center-surround posterior probability from the stimuli.
\par
For the case of mid-level features, we wanted to observe what structure in the stimuli would lead to center surround dependence. Note that for mid-level features, the notion of orientation is not as clear as in V1, where models may contain orientations in their filter parameterizations. 
\par
\begin{figure}[!]
   \centering
\includegraphics[]{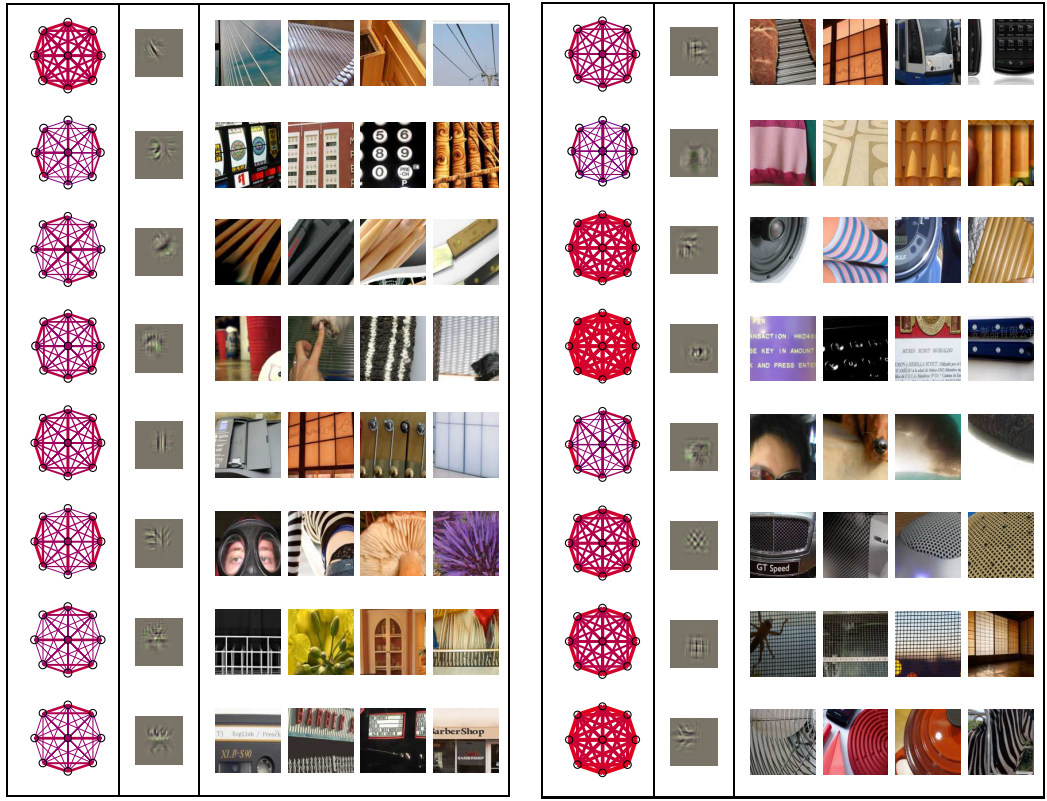}
  \caption{Tiling in Conv2 center-surround dependent units. Example units are ordered from lower learned prior of dependence (top unit on the left side table; .5193) to higher learned prior of dependence (bottom unit on the right side table; .9364). (First column) center-surround dependent covariances. Each black corresponds to the spatial location of the receptive fields. The line thickness between points depicts the strength of covariance between spatially shifted receptive fields. The size of the black circles depicts the variance. (Second column) Conv2 units visualization with a method adapted from \cite{MZeiler2014}. (Remaining columns) Image regions with high probability of being Center-Surround dependent and high activation values prior normalization according to our model. Note how regions can be seen as tiling the space with translations of the Conv2 unit receptive fields in directions with strong covariance.}\label{fig:tilings}
\end{figure}

Figure \ref{fig:tilings} shows some examples of image patches that cover the center-surround neighborhoods, for which the model finds a high posterior probability of Center-Surround dependence. Along with these images, a visualization of the receptive field of the second convolutional layer units is presented. In addition, The units depicted in Figure \ref{fig:tilings} are ordered based on the prior probability of center-surround dependence that is learned by our flexible normalization model. The top row of the figure corresponds to the lowest value of this prior probability among the units displayed in the figure.

\begin{figure}[t]
   \centering
\includegraphics[]{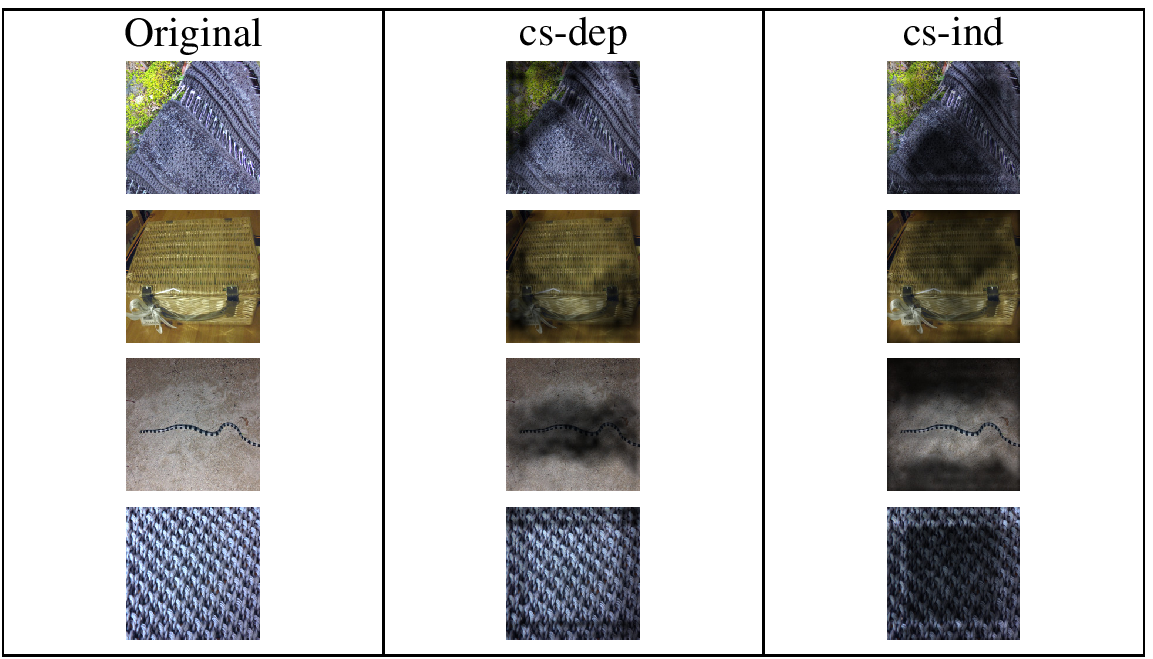}
  \caption{Inferred posterior distributions on ImageNet data. (left) Original images input to the CNN. For each image at each spatial location, we compute the geometric mean of posterior probabilities among all channels of the layer (Conv2). (middle) Areas that the model deemed as dependent while obscuring other areas. (right) Complementary display where high center-surround dependent areas are darker, instead.}\label{fig:spatial_posterior}
\end{figure}

As can be seen, for some of the Conv2 representations, the idea of co-linearity is present in the form of tiling of these mid-level features. The center surround covariance also captures this property. By looking at the receptive fields of the Conv2 units (second column from the left in Figure \ref{fig:tilings}) we can see that spatial arrangements of repetitions of these receptive fields seem  more natural in certain configurations. For instance, in the second row on the left table, vertical  arrangements of translated versions of the receptive field give a continued pattern as appreciated in the corresponding patches (right column of the table). Similarly, the seventh and eighth rows show horizontal preference. For other example units (e.g., the last 3 units which appear to capture texture-like structure), the covariance structure is more uniform across space.
\par
We also looked at the spatial distribution of the posterior probabilities for entire images from ImageNet. For each channel in the convolutional layer, we obtained the posterior probabilities and computed the geometric mean across channels. Figure \ref{fig:spatial_posterior} shows the relation between the image content and the posterior probability by shading areas with high (middle column) and low (right column) posterior probability of being center-surround dependent. As mentioned earlier, a high posterior probability of center-surround dependent activations can be an indicator of the homogeneity of the region under consideration. As we can see, the posterior probabilities from Conv2 units capture this homogeneity at a more structured level compared to previous work on V1-like filters, where homogeneous regions correspond to elongated lines in the preferred orientations of the filters.

\section{Discussion}

V1 normalization has been studied extensively, including flexible models of divisive normalization by the spatial context.
In modeling normalization of V1 units using a Gaussian Scale Mixture framework, previous work has shown that the learned covariance matrix (influencing the contextual normalization) has collinear structure \citep{RCoenCagli2012} (see also appendix). This reflects the oriented front end of V1 units and the collinear structure prevalent in images. Our work seeks to address richer receptive field structure beyond the level of V1-like units. We therefore build on a modified version of the model of \citep{RCoenCagli2012}, and use mid-level representations that are learned by convolutional networks trained for object recognition.
\par
The second convolutional layer of AlexNet combines responses from different units from the first layer into more elaborate features. However, it is not intuitively clear what is the covariance structure between such second layer units. The covariance structure can be understood as a template of what spatial arrangements lead to high center-surround dependence for a given model unit. Nor has there been emphasis on studying the statistical properties of more sophisticated units beyond V1-like oriented structure, such as those in the second layer of the CNN.
We therefore set out to study the statistics of the second layer of the CNN, and to examine the covariance structure that is learned. 
\par
First, we found that units in the Alexnet CNN had sparse marginal statistics and joint conditional statistical dependencies across space, similar to what has been observed for ICA and Gabor filters. Although this decreased in higher layers on average, the statistics in the second layer of AlexNet were still prominent. Further, we found in our simulations, that the learned covariances for Conv2 units do include more involved spatial arrangements, as well as collinearity. Images that were deemed as highly center-surround dependent for second layer units showed that the covariance structure can be attributed to tiling in more complex structures. This included, for instance, the boundary between two different textures. In addition, some tiling was more uniform across space, as for textures. Textures have also received attention in other studies in mid-level visual areas \citep{JFreeman2013,RRowekamp2017,NLaskar2018,CZiemba2018} and in generative models such as \citep{JPortilla2000,LGatys2015}.
\par
Furthermore, we found that from a technical perspective, adding independence among surround units in the center-surround independent component of the mixture \eqref{eq:binary_gsm} was crucial for learning the parameters of the model without imposing symmetry constraints. This is particularly necessary in the context of deep CNNs, and even more so at middle levels. Unlike Gabor filters or steerable pyramids, units of deep CNNs are not parametrized explicitly with respect to orientations, for instance. The same issue arises in cortical neurons, whereby higher neural areas beyond V1 combine orientations in a nonlinear manner, leading to rich structure that is not easily parameterized. 
\par
The model does a good job at reducing multiplicative dependencies, but it is not perfect. Currently, filters of the CNN are not learned jointly with flexible normalization. Another limitation of the model is that flexible normalization is incorporated in a single layer of the CNN. An important future direction which would require further technical development beyond the scope of this paper, is learning flexible normalization in multiple layers of the CNN (e.g., layers 1 and 2) simultaneously. Our approach can also be used with other classes of hierarchical architectures. 
\par
Previous work on flexible normalization has tested predictions for natural images in area V1 \citep{RCoenCagli2015}.
Our approach can be used to make predictions about when normalization might be relevant in higher visual cortical areas to reduce redundancy in non-trivial ways. In particular, we expect our model can make useful predictions for testing normalization of neural responses to large stimuli that extend beyond the classical receptive field in V2.
The inferred posterior probabilities expose spatial structure that matches well with some forms of homogeneity such as those present in textures. 
Our approach could also be used in the future to examine the implications for different classes of stimuli, such as naturalistic textures versus noise in V2 \citep{CZiemba2018}. Therefore, another pressing direction for future work is to test model predictions directly against cortical data in area V2.

\subsection*{Acknowledgments}
This work was kindly supported by the National Science Foundation (grant 1715475), and a hardware donation from NVIDIA.
\section*{Appendix}
\subsection{Maximum likelihood estimation of covariance}
Recall the pdf of our Gaussian scale mixture with Rayleigh mixer:
\begin{equation}\label{eq:pdf_of_gsm}
p_X(\mathbf{x}) = \frac{1}{\left(2 \pi \right)^{m/2} \vert\bm{\Lambda} \vert^{1/2}h^m} a^{1-m/2}K_{m/2-1}(a),
\end{equation}
where $K_{\lambda}(\cdot)$ is the modified Bessel function of the second kind, and
\begin{equation}\label{eq:GSM_normalization_gain}
a^2 = \frac{\mathbf{x}^{\mathrm{T}}\bm{\Lambda}^{-1}\mathbf{x}}{h^2}.
\end{equation}
Equation \eqref{eq:pdf_of_gsm} can be employed to compute the likelihood function of the covariance matrix $\bm{\Lambda}$. Furthermore, notice that the scale parameter $h$ can be simply dismissed by making part of the covariance\footnote{\normalsize{Adding scale back is straightforward.}}. If we take $\hat{\bm{\Lambda}}^2 = \bm{\Lambda}h^2$, \eqref{eq:pdf_of_gsm} becomes:
\begin{equation}\label{eq:pdf_of_gsm_noscale}
p_X(\mathbf{x}) = \frac{1}{\left(2 \pi \right)^{m/2} \vert\hat{\bm{\Lambda}} \vert^{1/2}} \hat{a}^{1-m/2}K_{m/2-1}(\hat{a}),
\end{equation}
where $\hat{a} = \mathbf{x}^{\mathrm{T}}\hat{\bm{\Lambda}}^{-1}\mathbf{x}.$ From this point on, to simplify notation, we will refer to $\hat{\bm{\Lambda}}$ as $\bm{\Lambda}.$ and $\hat{a}$ as simply $a$.

For an exemplar $\mathbf{x}_i$, the partial derivative of $\log$-likelihood function with respect to $\bm{\Lambda}^{-1}$ is given by:
\begin{eqnarray}
\nonumber \frac{\partial \log{L(\bm{\Lambda}\vert \mathbf{x}_i)}}{\partial \bm{\Lambda}^{-1}} = \frac{\partial \log{\left(p_X(\mathbf{x}_i)\right)}}{\partial \bm{\Lambda}^{-1}} & = & \frac{1}{p_X(\mathbf{x}_i)}\frac{\partial p_X(\mathbf{x}_i)}{\partial \bm{\Lambda}^{-1}}\\
 & = & \frac{\bm{\Lambda}}{2} - \frac{1}{2 a}\frac{K_{m/2}(a)}{K_{m/2-1}(a)}\mathbf{x}_i{\mathbf{x}_i}^{\mathrm{T}}.\label{eq:partial_derviative_log_likelihood}
\end{eqnarray}
Based on \eqref{eq:partial_derviative_log_likelihood}, we propose the following iterative update rule for $\bm{\Lambda}$
\begin{equation}\label{eq:fixed_point_covariance}
\bm{\Lambda}_{\textrm{new}} \leftarrow \frac{1}{N}\sum\limits_{i=1}^N g_{m}(a_i){\mathbf{x}_i}{\mathbf{x}_i}^{\mathrm{T}},
\end{equation}
where
\begin{equation}\label{eq:fixed_point_GSM_normalization_gain}
g_{m}(a_i) = \frac{1}{a_i}\frac{K_{m/2}(a_i)}{K_{m/2-1}(a_i)},
\end{equation}
and $a_i = \sqrt{{\mathbf{x}_i}^{\mathrm{T}}{\bm{\Lambda}_{\textrm{old}}}^{-1}\mathbf{x}_i}.$ 
\subsubsection*{Details about inference}
Here, we work out the inference procedure for the full covariance GSM. Let  $\setminus i$ denote the set of all indexes minus index $i$, and decompose the precision matrix $\bm{\Lambda}^{-1}$ into, 
\begin{equation}\label{eq:precision_matrix}
\bm{\Lambda}^{-1} = \left[ \begin{array}{cc}
(\bm{\Lambda}^{-1})_{\setminus i,\setminus i} & (\bm{\Lambda}^{-1})_{\setminus i, i} \\
(\bm{\Lambda}^{-1})_{i, \setminus i} & (\bm{\Lambda}^{-1})_{i, i} \\
\end{array}\right]. 
\end{equation} 
It can be shown that 
\begin{equation}\label{eq:unreduced_conditional_x_given_c}
\EV{G_i|\mathbf{x}} =  \frac{c_i}{\sqrt{a_{\setminus i}}}\left(\frac{a}{a_{\setminus i}}\right)^{\frac{m}{2} - 1}\frac{\vert\bm{\Lambda}\vert^{\frac{1}{2}}}{\sigma_{i}\vert((\bm{\Lambda}^{-1})_{\setminus i,\setminus i})^{-1}\vert^{\frac{1}{2}}}\frac{K_{\frac{m-1}{2}}(a_{\setminus i})}{K_{\frac{m}{2}-1}(a)},
\end{equation}
where ${\sigma_{i}}^{2} = (\bm{\Lambda})_{i,i}$, and
\begin{equation}\label{eq:energy_of_surround}
\begin{split}
{a_{\setminus i}}^2 = \:& {\mathbf{x}_{\setminus i}}^{\mathrm{T}}(\bm{\Lambda}^{-1})_{\setminus i,\setminus i}\mathbf{x}_{\setminus i} +
 					  2 x_{i} {\mathbf{x}_{\setminus i}}^{\mathrm{T}}(\bm{\Lambda}^{-1})_{\setminus i,i} + \\
 					  & + {x_{i}}^{2} \left[(\bm{\Lambda}^{-1})_{i, \setminus i}((\bm{\Lambda}^{-1})_{\setminus i,\setminus i})^{-1}		  (\bm{\Lambda}^{-1})_{\setminus i,i} + {\sigma_{i}}^{-2} \right].  					  
\end{split}
\end{equation}
Noticing that 
\begin{equation}\label{eq:submatrix_inv_identity}
(\bm{\Lambda}^{-1})_{i,i} = \sigma{i}^{-2} + (\bm{\Lambda}^{-1})_{i, \setminus i}((\bm{\Lambda}^{-1})_{\setminus i,\setminus i})^{-1}		  (\bm{\Lambda}^{-1})_{\setminus i,i}, 
\end{equation} 
yields $a_{\setminus i} = a$. Furthermore, since  $\sigma_{i}^2 = \vert\bm{\Lambda}\vert \vert(\bm{\Lambda}^{-1})_{\setminus i,\setminus i}\vert$,
\begin{equation}\label{eq:reduced_conditional_x_given_c}
\EV{G_i|\mathbf{x}} =  \frac{x_i}{\sqrt{a}}\frac{K_{\frac{m-1}{2}}(a)}{K_{\frac{m}{2}-1}(a)}.
\end{equation}

\subsection{Mixture of Gaussian scale mixtures}\label{sec:mixture_GSM}
The mixture model has the following general form:
\begin{equation}\label{eq:mixture_model}
p_X(\mathbf{x}) = \sum\limits_{\alpha \in \mathcal{A}} \Pi_{\alpha}p_{X}(\mathbf{x}\vert \bm{\Lambda}_{\alpha})
\end{equation}
Parameter estimation for the above model \eqref{eq:mixture_model} can be solved using expectation maximization (EM) algorithm. In particular, we use the conditional EM algorithm to update the parameters of each of the mixture components. For each partial \textbf{E-step}, we compute the posterior distributions over the assignment variable:
\begin{equation}\label{eq:posterior_E_step}
q(\alpha, \mathbf{x}_i) = \frac{\Pi_{\alpha}p_X(\mathbf{x}_i\vert\bm{\Lambda}_{\alpha})}{\sum\limits_{\alpha' \in \mathcal{A}}\Pi_{\alpha'}p_X(\mathbf{x}_i\vert\bm{\Lambda}_{\alpha'})},\:\textrm{for all}\:\alpha\in\mathcal{A}.
\end{equation}
In the partial \textbf{M-step}, we update all the mixture probabilities using \eqref{eq:posterior_E_step},
\begin{equation}\label{eq:mix_prob_M_step}
\Pi_{\alpha'} \leftarrow \frac{1}{N} \sum\limits_{i=1}^Nq(\alpha', \mathbf{x}_i),\:\textrm{for all}\:\alpha'\in\mathcal{A}, 
\end{equation}
and the corresponding covariance $\bm{\Lambda}_{\alpha}$ using a modified version of the fixed-point \eqref{eq:fixed_point_covariance}, as follows:
\begin{equation}\label{eq:mgsm_fixed_point_covariance}
\bm{\Lambda}_{\alpha} \leftarrow \frac{\sum\limits_{i=1}^N q(\alpha, \mathbf{x}_i)g_{m}(\mathbf{x}_i\vert \bm{\Lambda}_{\alpha}){\mathbf{x}_i}{\mathbf{x}_i}^{\mathrm{T}}}{\sum\limits_{j=1}^N q(\alpha, \mathbf{x}_j)}.
\end{equation}
where $g_{m}(\mathbf{x}_i\vert \bm{\Lambda}_{\alpha}) = g_{m}(\sqrt{{\mathbf{x}_i}^{\mathrm{T}}{\bm{\Lambda}_{\alpha}}^{-1}\mathbf{x}_i}) $ from \eqref{eq:fixed_point_GSM_normalization_gain}
We use a single fixed-point iteration per partial CEM iteration. The proposed fixed point update increases the likelihood at each iteration.
\subsubsection*{Two-GSM mixture model for flexible normalization}
Gaussian scale mixture models have been used to explain non-linear dependencies among linear decompositions of natural stimuli such as images. In the simplest case, it is assumed that such dependencies carry over the entire stimuli. For example, in vision, commonly used approaches of local contrast normalization apply the same normalization scheme across the entire image. Spatial pools for normalization have been applied to explain responses to redundant stimuli. While this model is able to account for suppressions of unit responses where spatial context is redundant, it can also lead to suppression in cases where context may not be redundant. A flexible normalization that suppresses responses only when the spatial context is deemed as redundant can be constructed as a mixture of GSMs. A simple version considers a component with full center-surround dependencies, a second component representing the center-surround independence results from the product of a center only and surround only GSM distributions:
\begin{equation}\label{eq:binary_gsm}
p_X(\mathbf{x}) = \Pi_{\textrm{cs}}p_X(\mathbf{x}\vert \bm{\Lambda}_{\textrm{cs}}) + (1 - \Pi_{\textrm{cs}})p_{X_{\textrm{c}}}(\mathbf{x}_{\textrm{c}}\vert \bm{\Lambda}_{\textrm{c}})p_{X_{\textrm{s}}}(\mathbf{x}_{\textrm{s}}\vert \bm{\Lambda}_{\textrm{s}}),
\end{equation}  
where $\mathbf{x}_{\textrm{c}}$ and $\mathbf{x}_{\textrm{s}}$ denote the sub-vectors of $\mathbf{x}$ containing the center and surround variables, respectively.
The variants of the EM steps prsented in \eqref{eq:posterior_E_step}, \eqref{eq:mix_prob_M_step}, and \eqref{eq:mgsm_fixed_point_covariance} are discussed below. For each partial \textbf{E-step}   
\begin{eqnarray}
q(\textrm{cs}, \mathbf{x}_i) & = & \frac{\Pi_{\textrm{cs}}p_X(\mathbf{x}_i\vert\bm{\Lambda}_{\textrm{cs}})}{Q(\mathbf{x}_i)}, \\ 
q(\textrm{cs}\independent, \mathbf{x}_i) & = & \frac{(1 - \Pi_{\textrm{cs}})p_{X_{\textrm{x}}}(\mathbf{x}_{i,\textrm{c}}\vert \bm{\Lambda}_{\textrm{c}})p_{X_{\textrm{s}}}(\mathbf{x}_{i,\textrm{s}}\vert \bm{\Lambda}_{\textrm{s}})}{Q(\mathbf{x}_i)}  = 1 - q(\textrm{cs}, \mathbf{x}_i) \\
Q(\mathbf{x}_i) & = &  \Pi_{\textrm{cs}}p_X(\mathbf{x}_i\vert\bm{\Lambda}_{\textrm{cs}}) + (1 - \Pi_{\textrm{cs}})p_{X_{\textrm{c}}}(\mathbf{x}_{i,\textrm{c}}\vert \bm{\Lambda}_{\textrm{c}})p_{X_{\textrm{s}}}(\mathbf{x}_{i,\textrm{s}}\vert  \bm{\Lambda}_{\textrm{s}})\\
\end{eqnarray}
Each partial \textbf{M-step} updates the center-surround dependent probability using \eqref{eq:posterior_E_step},
\begin{equation}\label{eq:binary_mix_prob_M_step}
\Pi_{\textrm{cs}} \leftarrow \frac{1}{N} \sum\limits_{i=1}^Nq(\textrm{cs}, \mathbf{x}_i). 
\end{equation}
Three partial \textbf{M-step} updates are required:
\begin{enumerate}
\item A center-surround dependent covariance $\bm{\Lambda}_{\textrm{cs}}$ update, 
\begin{equation}\label{eq:binary_mgsm_fixed_point_cs_covariance}
\bm{\Lambda}_{\textrm{c}} \leftarrow \frac{\sum\limits_{i=1}^N q(\textrm{cs}, \mathbf{x}_i)g_{m}(\mathbf{x}_i\vert \bm{\Lambda}_{\textrm{cs}}){\mathbf{x}_i}{\mathbf{x}_i}^{\mathrm{T}}}{\sum\limits_{j=1}^N q(\textrm{cs}, \mathbf{x}_j)},
\end{equation}
\item a center-only covariance
\begin{equation}\label{eq:binary_mgsm_fixed_point_c_covariance}
\bm{\Lambda}_{\textrm{c}} \leftarrow \frac{\sum\limits_{i=1}^N (1 - q(\textrm{cs}, \mathbf{x}_{i}))g_{m_{\textrm{c}}}(\mathbf{x}_{i, \textrm{c}}\vert \bm{\Lambda}_{\textrm{c}}){\mathbf{x}_{i, \textrm{c}}}{\mathbf{x}_{i, \textrm{c}}}^{\mathrm{T}}}{\sum\limits_{j=1}^N 1 - q(\textrm{cs}, \mathbf{x}_j)}.
\end{equation}
\item and a surround-only covariance
\begin{equation}\label{eq:binary_mgsm_fixed_point_s_covariance}
\bm{\Lambda}_{\textrm{s}} \leftarrow \frac{\sum\limits_{i=1}^N (1 - q(\textrm{cs}, \mathbf{x}_{i}))g_{m_{\textrm{s}}}(\mathbf{x}_{i, \textrm{s}}\vert \bm{\Lambda}_{\textrm{s}}){\mathbf{x}_{i, \textrm{s}}}{\mathbf{x}_{i, \textrm{s}}}^{\mathrm{T}}}{\sum\limits_{j=1}^N 1 - q(\textrm{cs}, \mathbf{x}_j)}.
\end{equation}
\end{enumerate}
\noindent\textbf{Re-parameterization:}\\
\indent To simplify computations and directly enforce the nonnegative definiteness in our covariance estimation, we re-parametrize the likelihood function. Let us write, 
\begin{equation}\label{eq:reparametrization}
\bm{\Lambda} = \mathbf{B}^{\mathrm{T}}\mathbf{B},\:\textrm{and}\: \bm{\Lambda}^{-1} = \mathbf{A}^{\mathrm{T}}\mathbf{A}.
\end{equation}
Then, $\mathbf{B} = \mathbf{A}^{-\mathrm{T}}$ and 
\begin{equation}\label{eq:partial_derivative_log_likelihood_reparam}
\frac{\partial \log{L(\bm{\Lambda}\vert \mathbf{x}_i)}}{\partial \mathbf{A}} =  \mathbf{B} - \frac{1}{a}\frac{K_{m/2}(a)}{K_{m/2-1}(a)}\mathbf{A}\mathbf{x}_i{\mathbf{x}_i}^{\mathrm{T}},
\end{equation}
which yields the following fixed point update:
\begin{equation}\label{eq:fixed_point_constrained_covariance}
\mathbf{B}_{\textrm{new}} \leftarrow \frac{1}{N}\sum\limits_{i=1}^N g_{m}(a_i){\mathbf{A}_{\textrm{old}}\mathbf{x}_i}{\mathbf{x}_i}^{\mathrm{T}} \end{equation}

\subsubsection*{A center-surround independent model with independent surround units}
In this model the center surround independent component has the extra property that requires surround units to be independent from each other. One consequence of this requirement is that the surround covariance $\bm{\Lambda}_{\textrm{s}}$ becomes a diagonal matrix. Note that diagonal covariance is a necessary but not sufficient condition for independence in this case. The main feature for independence is that each one of the surround units has its own mixer (scaling rather than mixing) variable instead of a shared mixer as it is the case in the model previously discussed. If the mixer (scaling) variables are Rayleigh distributed, each surround unit $\ell$ in the center-surround independent component has a Laplace distribution:
\begin{equation}\label{eq:laplace_distribution}
f_{\ell}(x) = \frac{1}{2 \sqrt{\left(\bm{\Lambda}_{\textrm{c}}\right)_{\ell, \ell}}}\exp{\left(-\frac{\vert x \vert}{\sqrt{\left(\bm{\Lambda}_{\textrm{c}}\right)_{\ell, \ell}}}\right)},  
\end{equation}
where $\left(\bm{\Lambda}_{\textrm{c}}\right)_{\ell, \ell}$ denotes the diagonal element of the surround covariance matrix $\bm{\Lambda}_{\textrm{c}}$. Note that this matrix has zero off-diagonal elements by definition. In this model, 
\begin{equation}\label{eq:BGSM_surround_independent}
p_{X_{\textrm{s}}}(\mathbf{x}_{i,\textrm{s}}\vert \bm{\Lambda}_{\textrm{s}}) = \prod\limits_{\ell \in \mathcal{S}}f_{\ell}\left(\left(\mathbf{x}_{i,\textrm{s}}\right)_{\ell}\right)
\end{equation}
In this modified version \eqref{eq:binary_mgsm_fixed_point_s_covariance} becomes:
\begin{equation}\label{eq:binary_mgsm_fixed_point_s_indp_covariance}
\left(\bm{\Lambda}_{\textrm{s}}\right)_{\ell,\ell} \leftarrow \frac{\sum\limits_{i=1}^N (1 - q(\textrm{cs}, \mathbf{x}_{i}))\left\vert \left({\mathbf{x}_{i, \textrm{s}}}\right)_{\ell} \right\vert}{\sum\limits_{j=1}^N 1 - q(\textrm{cs}, \mathbf{x}_j)}\sqrt{\left(\bm{\Lambda}_{\textrm{s}}\right)_{\ell,\ell}}.
\end{equation}
The rest of the \textbf{EM} algorithm proceeds in the same way as described in \eqref{eq:binary_mgsm_fixed_point_cs_covariance} and \eqref{eq:binary_mgsm_fixed_point_c_covariance}.

\subsubsection*{Matching covariances for inference}
Here, we describe how we obtain the transformation $\mathbf{Q}$ for equation (7) in the paper.
Assuming that both matrices are full rank, we can write $\bm{\Lambda}_{\textrm{cs}} = \mathbf{A}^{\mathrm{T}}\mathbf{A}$ and  $\bm{\Lambda}_{\textrm{cs}\independent} = \mathbf{B}^{\mathrm{T}}\mathbf{B}$. Furthermore, there exists a transformation $\mathbf{Q}$ such that:
\begin{equation}\label{eq:covariance_transformation}
\mathbf{Q}^{\mathrm{T}}\bm{\Lambda}_{\textrm{cs}\independent}\mathbf{Q} = \bm{\Lambda}_{\textrm{cs}}, 
\end{equation}
which is simply given by $\mathbf{Q} = \mathbf{B}^{-1}\mathbf{A}$. 

\subsection{Judging the effectiveness of normalization}
As noted above, the Gaussian scale mixture introduces a multiplicative coupling between variables that cannot be removed by linear means. This coupling is captured by a simple dependency measure based on the energy of the variables. For zero mean, unit variance, and mutually independent $G_i$ and $G_j$, define $X_i = C_iV$ and $X_j = G_jV$ where mixer $V$ is also independent of $G_i$ and $G_j$. The covariance of $X_i^2$ and $X_j^2$ is given by
\begin{eqnarray}
\nonumber \EV{(X_i^2 - \EV{X_i^2})(X_j^2 - \EV{X_j^2})} & = & \EV{X_i^2X_j^2} - \EV{X_i^2}\EV{X_j^2} \\
\nonumber											     & = & \EV{G_i^2V^2G_j^2V^2} - \EV{G_i^2V^2}\EV{G_j^2V^2} \\
\label{eq:energy_covariance}					         & = & \EV{V^4} - \EV{V^2}^2.
\end{eqnarray}
The strength of the coupling depends on the spread of $V$. A perfect inversion of the coupling, which would require explicit values of $V$, would make the expression \eqref{eq:energy_covariance} zero. Here, we use a related measure, the correlation between squared responses \citep{ACoates2011}. This measure has been used to select groups of receptive fields that should be processed together in a subsequent layer of a deep network. The correlation between squared responses  is computed in a two-step process. First, variables $X$ are decorrelated by whitening using ZCA. For the pair whitened variables $(\tilde{X}_i, \tilde{X}_j)$, correlation of squared responses is given by:
\begin{equation}\label{eq:squared_correlation}
S(\tilde{X}_i, \tilde{X}_j) = \frac{\EV{(\tilde{X}_i^2\tilde{X}_j^2 - 1)}}{\sqrt{\EV{(\tilde{X}_i^4 - 1)}\EV{(\tilde{X}_j^4 - 1)}}}
\end{equation}

\subsection{Additional simulations}
\subsubsection*{Flexible normalization on the first convolutional layer of AlexNet}
We also trained our flexible normalization model on the responses of the first convolutional layer of AlexNet. The filters on this layer resemble the patterns that have been identified to elicit vigorous responses in V1 neurons. This is not the first time the flexible normalization model has been applied to filter modeling V1. For previous work, we refer the readers to \cite{RCoenCagli2009, RCoenCagli2012, RCoenCagli2015}. Nevertheless, to the best of our knowledge this is the first time the flexible normalization model has been applied to filters learned from data in a supervised learning task. 
In previous work, the orientation of the filters, which was known, was employed to restrict the model fitting by adding symmetry constraints to the covariance matrices of the model. As we explained in the main text, our modified model does not employ these symmetry constraints, but forces the surround variables to be fully independent, which translates into having a separate mixer variable for each one of them.  
\subsubsection*{Covariance structure of the surround components of the first convolutional AlexNet}
\begin{figure}[!th]
   \centering
\includegraphics[]{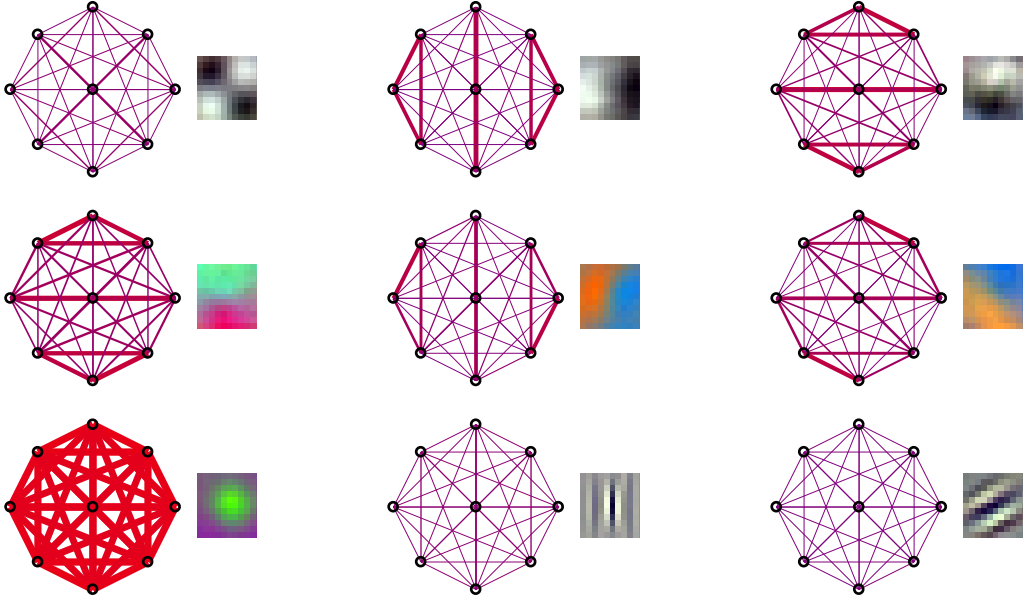}
\caption{Covariance structure for different units in the first convolutional layer of AlexNet. We display the covariance structure of the surround pool along with the visualization of the corresponding filter. Thicker lines mean larger magnitude of the correlation. Line color linearly interpolates from blue for negative values to red for positive values}\label{fig:covariance_layer1}
\end{figure}

Similarly to Figure 4a in the paper, we visualize the covariance structure of the surround covariance (Figure \ref{fig:covariance_layer1}). As we can see, low frequency filters expose stronger correlation in their responses than the high frequency filters. Also the orientation of the filter is reflected in the covariance structure of the model, similar to the results obtained in \citep{RCoenCagli2009} for wavelet filters.
\subsubsection*{High order correlations between surround components of the first convolutional layer of AlexNet}
\begin{figure}
\centering
\includegraphics[trim=110 0 50 0,clip,height=0.5\linewidth]{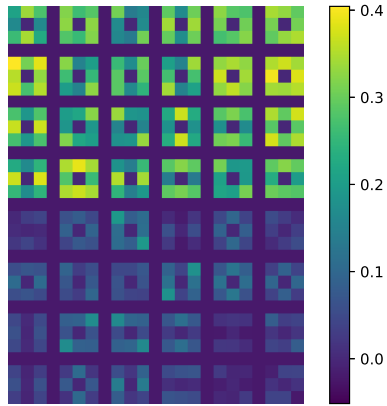}
\caption{Correlation of energies between center and surround responses for a subset of units of the first convolutional layer of AlexNet. The upper half correspond to the correlation before normalization and the bottom half after flexible normalization.}\label{fig:cs_corr_all_conv1}
\end{figure}

Here, we show the correlation of energies for the first layer units of AlexNet before and after normalization (Figure \ref{fig:cs_corr_all_conv1}). We can see that the normalization procedure reduces the energy correlation significantly.
In addition to the squared correlation, we also visualize the normalized conditional histograms before and after normalization, as well as the marginal distributions of the center variable.
\begin{figure}[!t]
   \centering
\includegraphics[]{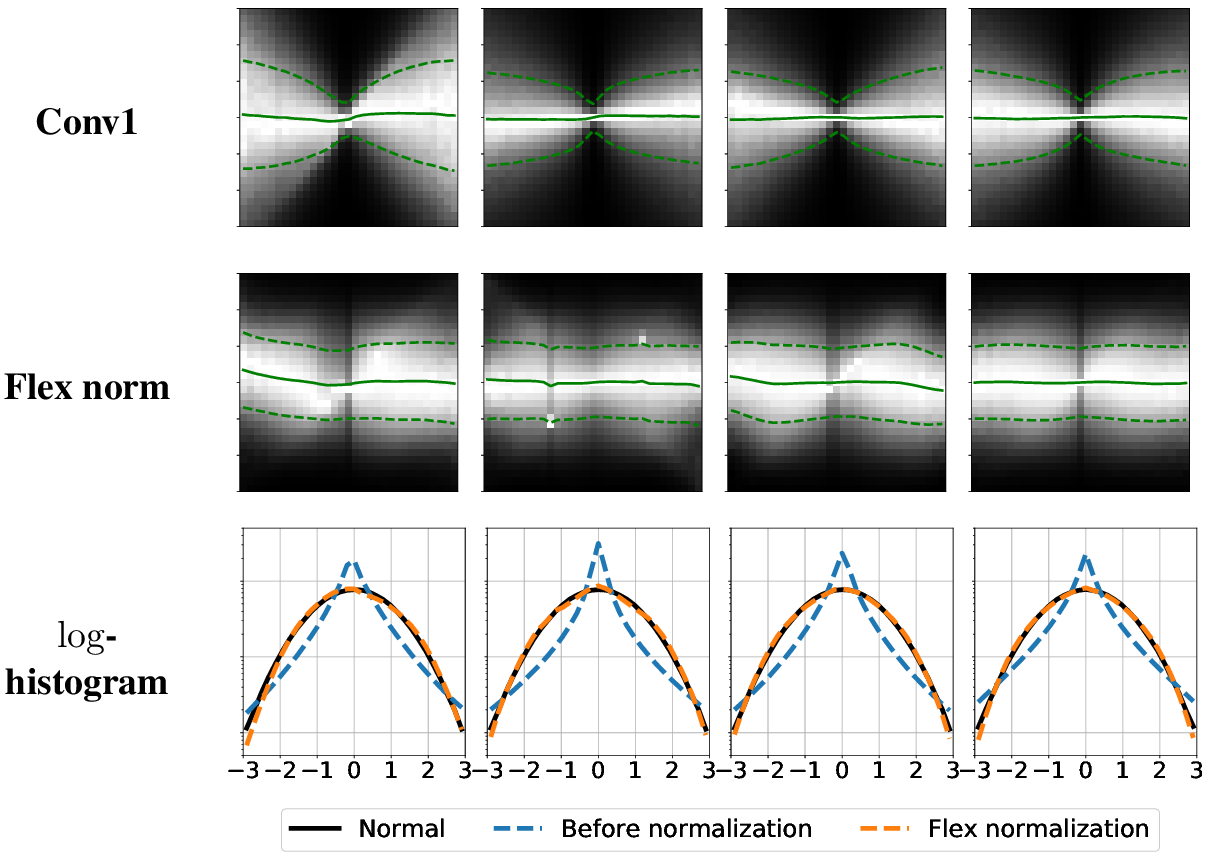}
  \caption{Normalized conditional histograms between center and surround responses from the first convolutional layer of AlexNet. The first two rows from the top are the conditional distributions before and after flexible normalization. The third row are the corresponding $\log$ histograms.}\label{fig:bowties_log_histograms_conv1}
\end{figure}
\subsubsection*{Single GSM normalization of second layer units of AlexNet}
In addition to flexible normalization, we looked at a simpler model which assumes the coupling between center and surround units remains the same across the entire image. This model is a particular case of the flexible normalization where $\Pi_{\textrm{cs}} = 1$. In this model, only the center surround dependent covariance $\bm{\Lambda}_{\textrm{cs}}$ is of interest. As shown in more detail for the population statistics in the main text,  the single GSM model reduces dependencies and makes the marginal distributions closer to Gaussian, but not as much as the mixture of GSMs model.
\begin{figure}[!t]
   \centering
  \includegraphics[]{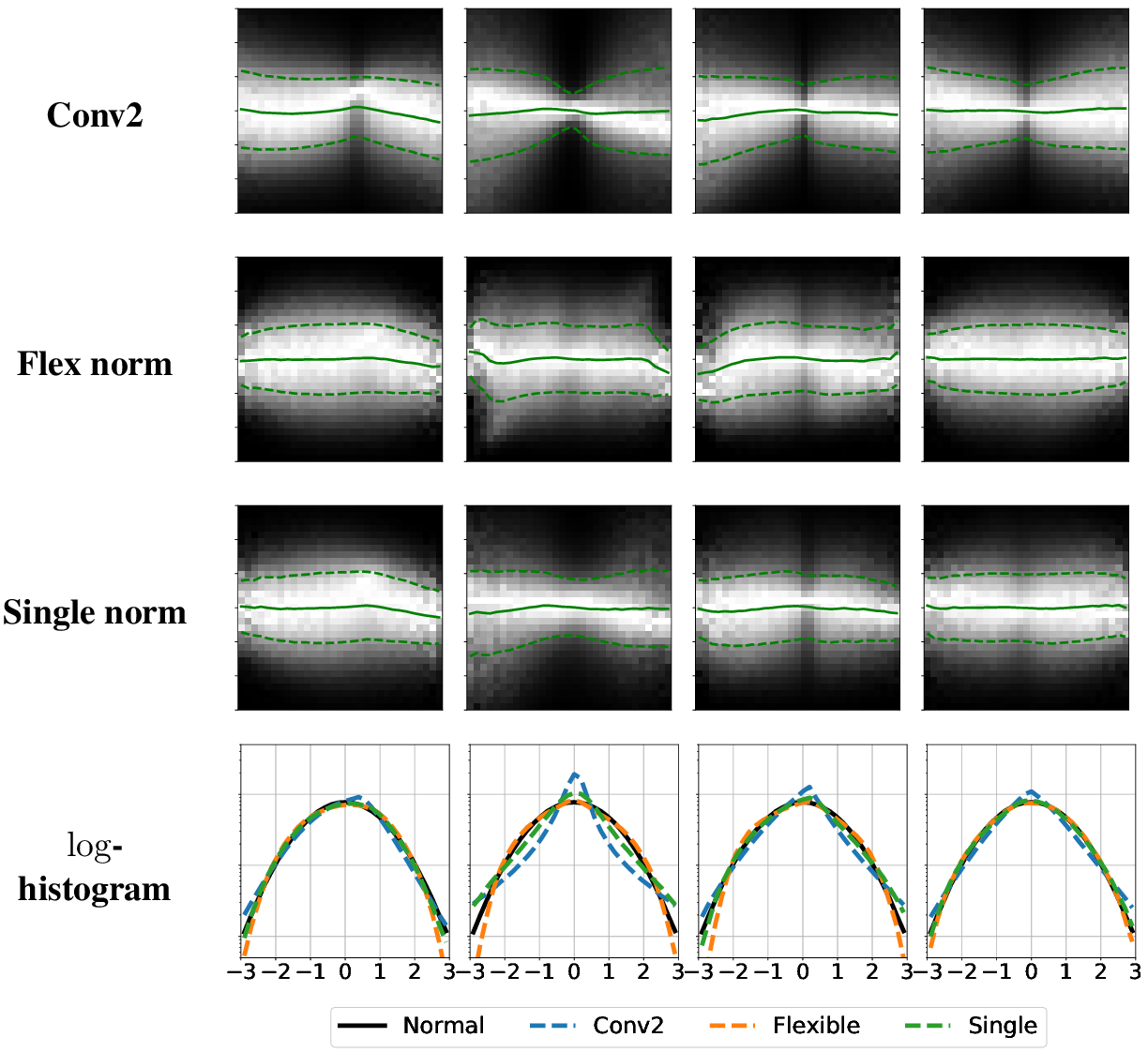}
  \caption{Comparison between flexible and single GSM normalized conditional histograms between center and surround responses from the second convolutional layer of AlexNet. First three rows from the top show the conditional distributions before and after flexible normalization and single GSM normalization. The fourth row shows the corresponding $\log$-histograms.}\label{fig:bowties_log_histograms_cano_conv2}
\end{figure}

\bibliographystyle{apa}
\bibliography{../bibliography/computer_vision,../bibliography/comp_neuroscience,../bibliography/machine_learning,../bibliography/statistics,../bibliography/signal_processing}

\end{document}